\documentclass[twoside]{article}
\usepackage{epsfig,fleqn,espcrc2}

\usepackage{graphicx}
\usepackage[figuresright]{rotating}

\newcommand{\AmS}{{\protect\the\textfont2
  A\kern-.1667em\lower.5ex\hbox{M}\kern-.125emS}}

\def\efbsofb{f_{B_s}/f_B}

\def\fbsofb{$f_{B_s}/f_B$}

\def\gtwid{\raise.3ex\hbox{$>$\kern-.75em\lower1ex\hbox{$\sim$}}}
\def\ltwid{\raise.3ex\hbox{$<$\kern-.75em\lower1ex\hbox{$\sim$}}}

\def\et{{\it et al.}}

\def\prl#1{Phys.\ Rev.\ Lett.\ {\bf #1}}
\def\prd#1{Phys.\ Rev.\ {\bf D#1}}

\def\npb#1{Nucl.\ Phys.\ {\bf B#1}}

\def\bangalore{Nucl.\ Phys.\ {\bf B} (Proc.\ Suppl.) {\bf 94} (2001)}

\title{Heavy-light decay constants with three dynamical flavors}

\author{ C.~Bernard,\hskip-0.03in
\address{{\vskip-0.10in{\hskip 0.07in Department of Physics, Washington
University, St.~Louis, MO 63130, USA}}} % "a"
\hskip-0.03in\thanks{presented by C.\ Bernard}
T.~Burch,\hskip-0.03in
\address{Department of Physics, University of Arizona, Tucson, AZ 85721, USA} %""b"
S.\ Datta,\hskip-0.03in
\address{Department of Physics, Indiana University, Bloomington, IN 47405, USA} % "c"
T.~DeGrand,\hskip-0.03in
\address{Physics Department, University of Colorado, Boulder, CO 80309, USA} %
C.~DeTar,\hskip-0.03in
\address{Physics Department, University of Utah, Salt Lake City, UT 84112, USA}
Steven~Gottlieb,$\!\null^{\rm c}$ %c
\advance\baselineskip -2pt
Urs~M.~Heller,\hskip-0.03in
\address{SCRI, Florida State University, Tallahassee, FL 32306-4130, USA} %"f"
K.~Orginos,\hskip-0.03in
\address{RIKEN BNL Research Center, Upton, New York 11973, USA} %""g"
R.~Sugar\hskip0.005in
\address{Department of Physics, University of California, Santa Barbara, CA
93106, USA} %"h"
and
D.~Toussaint$\null^{\rm b}$ %b
\advance\baselineskip -2pt
} %end \author
       
\begin{document}

\begin{abstract}
We present preliminary results for the heavy-light leptonic decay
constants in the presence of three light dynamical flavors.  
We generate dynamical configurations with improved staggered
and gauge actions and analyze them for heavy-light physics with
tadpole improved clover valence quarks.
%The configurations span a wide range of dynamical quark mass and
%are matched in lattice spacing at $\approx\! 0.13$ fm, as determined by the
%static quark potential. We have additional data with two
%flavors and with quenched lattices for comparison, as well as quenched
%data on a finer lattice: $a\approx 0.09$ fm.  
When the scale is set by $m_\rho$, we find an increase of
$\approx\!23\%$
in $f_B$ with three dynamical flavors over the quenched case.
%When the scale is instead set by the potential,
%the three-flavor result
%for $f_B$ extrapolated to physical quark mass is unchanged within errors;
%this is not true in the quenched approximation.
Discretization errors appear to be small ($\ltwid3\%$) in the
quenched case but have not yet been measured in the dynamical case.
\vspace{-1pc}
\end{abstract}

\maketitle

The computation of leptonic decay constants for heavy-light mesons plays
a key role in the extraction of CKM matrix elements from experiment and has been
a focus of lattice QCD calculations for many years.  Here, we report
on the first attempt to compute such decay constants in the presence
of $N_F\!=\!3$ light sea quarks ($u$, $d$, $s$).
For the generation of the dynamical lattices,
we use the ``Asqtad'' action \cite{IMP_KS}, which consists of
order $\alpha_s a^2, a^4$ improved staggered quarks and
order $\alpha_s^2 a^2, a^4$ Symanzik improved glue \cite{IMP_GAUGE,ALFORD}.
This action has small discretization
errors for many  quantities \cite{IMP_SCALING,MILC_POTENTIAL}.
%Because heavy staggered quarks on the lattice are problematic, 
For the valence quarks,
we employ a tadpole-improved (Landau link) clover action and
the Fermilab formalism \cite{EKM}.  The analysis is done in a ``partially
quenched'' %\cite{PARTIAL_QUENCHING}
manner: 
the valence quark masses are extrapolated to their 
physical values with the sea quark masses held fixed.
The chiral extrapolation of the sea quark masses (at fixed lattice spacing)
is performed afterwards.

Table \ref{tab:lattices} shows the parameters of our lattices and the current
state of this project.  As we decrease the dynamical quark masses from 
large values in the three-flavor case, we keep the three 
masses degenerate until the physical strange quark
mass is reached.  We then keep the strange quark mass fixed as we further decrease 
$m_{u,d}$. 

Configurations with varying dynamical quark mass (including quenched and
two-flavor configurations) have been matched in lattice spacing using
the quantity $r_1$ \cite{MILC_POTENTIAL}, which is defined from the static quark
potential at shorter distance than $r_0$ \cite{SOMMER}.
From the value $r_0\approx 0.5$ fm  
we have found $r_1\approx 0.35$ fm \cite{MILC_SPECTRUM} in full QCD.
Note that the errors in these phenomenological values may be as large
as $10\%$ \cite{SOMMER}; we therefore prefer
to use $m_\rho$ 
in setting the absolute scale.
Details of the lattice generation and
light quark results are in Ref.~\cite{MILC_SPECTRUM}.
Our analysis of heavy-light decay constants follows Refs.~\cite{MILC_PRL,MILC_LAT00}; 
we describe only the important differences below.

%it must be kept in mind 
%that the physical values of $r_0$ and $r_1$ have perhaps 5 to 10\%
%error \cite{SOMMER} because they are not directly measurable in experiment.  
%Thus, 
%The clover action for the valence quarks is tree level tadpole improved \cite{LEPMAC}
%using the Landau gauge link.

\def\tr{\phantom{3}}
\begin{table}[htb]
\caption{Status of heavy-light running. We show the nominal value of $a$
determined from $r_1=0.35$ fm.  All 0.13 fm lattices are $20^3\times64$;
all 0.09 fm lattices are  $28^3\times96$.  ``\# gen.''\ is the number of
configurations that have been generated; ``\# $f_B$''\ is the number
on which $f_B$ has been calculated.
}
\vspace{0.05in}
\begin{center}
\begin{tabular}{ccccc}
   dynamical & $\beta$ & $a$ (fm) & \#    & \#  \\
              $am_{u,d}/am_s$ & &      & gen.   & $f_B$  \\
\hline
\hline
    $\infty$/$\infty$ & 8.00 & 0.13 &    408 &  290  \\
\hline
    0.02/$\infty$ & 7.20 & 0.13 &   536 &  411  \\
\hline
    0.40/0.40   & 7.35 & 0.13 &   332 &   --  \\
    0.20/0.20   & 7.15 & 0.13 &   341 &  341  \\
    0.10/0.10   & 6.96 & 0.13 &   339 &  --  \\
    0.05/0.05   & 6.85 & 0.13 &   425 &  425  \\
    0.04/0.05   & 6.83 & 0.13 &   351 &   --  \\
    0.03/0.05   & 6.81 & 0.13 &   564 &  193  \\
    0.02/0.05   & 6.79 & 0.13 &   486 &  486  \\
    0.01/0.05   & 6.76 & 0.13 &   407 &  399  \\
\hline
\hline
    $\infty$/$\infty$ & 8.40 & 0.09 &    417 &  200  \\
\hline
    0.031/0.031   & 7.18 & 0.09 &   162 &  40  \\
    0.0124/0.031   & 7.11 & 0.09 &   25 &  --  \\
\hline
\hline
\end{tabular}
\label{tab:lattices} 
\end{center}
\vskip -1.2truecm
\end{table} 

On all lattice sets, we compute quark propagators for 5 light and 5 heavy masses.
This gives good control over both chiral 
and heavy quark interpolations/extrapolations. 
In Ref.~\cite{MILC_PRL}, with 3 light quark masses,
the chiral extrapolation was a major source of systematic error.  Here, changing
from linear to quadratic chiral fits of decay constants changes the results
by $\ltwid2\%$.

The (tadpole improved) perturbative renormalization of the heavy-light
axial current, $Z_A^{\rm tad}$, has not yet been calculated.  At present,
we therefore proceed as follows at $a\!\approx\!0.13$ fm:

(1) Define boosted coupling $\alpha^P_s$ from the plaquette following
Refs.~\cite{ALFORD,WEISZ}, and then
define the 1-loop coefficient $\zeta_A$ 
by $Z_A^{\rm tad} = 1 + \alpha_s^P \zeta_A$.
%\begin{equation}
%\alpha_s^P = -{\ln({1\over3}{\rm Re\; Tr} <U_{plaq}>) \over 3.06839}\ . 
%\end{equation}

(2) Fix the scale from $m_\rho$ on each set,
with valence quarks extrapolated to physical values.

(3) Fix $\zeta_A$ by demanding 
 that $f_B^{\rm quench}=169$ MeV  (the MILC 
continuum-extrapolated result \cite{EXPLAINfB}).
This gives $Z_A^{\rm tad}$ on all the 0.13 fm lattices,
since $\zeta_A$ is independent of the number of dynamical flavors. 

Note that $\zeta_A$ is dependent on the heavy quark
mass through the dimensionless quantity $am$.  To test scaling 
%with our quenched lattices at $a\approx 0.09$ fm, 
we first
repeat the above procedure for $f_D$ at  0.13 fm  
and then interpolate to the correct
quark mass for $f_B$ at 0.09 fm. There is a 2\% change in $\zeta_A$ for $f_B$
between the two lattice spacings.  Because this change results in only a 0.3\% effect 
on $f_B$, we believe that systematic effects in the %scaling test
procedure are negligible.

Note also that comparisons of $f_B$ with $f_B^{\rm quench}$ or changes
of $f_B$ with $a$ are only
very weakly dependent on our assumption about the  continuum value of
$f_B^{\rm quench}$.  Further, the ratio $f_{B_s}/f_B$ is essentially
independent of perturbation theory.
%except for $\cO(am_q)$ terms, where $m_q$ is a light quark mass.
%However, the absolute magnitude of  $f_B$ is strongly dependent on
%the assumption for $f_B^{\rm quench}$.

In the plots below, we show two sets of errors.  The smaller ones
represent the statistical errors from the fits, computed
by jackknife.  We use the same fitting ranges in time for all lattices with 
the same spacing.  The larger error bars take into account variations caused
by changing the fitting ranges while keeping confidence levels  and statistical
errors reasonable.  
%On some sets (notably the $N_F\!=\!2$ set),
%long range ``wiggles'' in the heavy-light plateaus make these second errors
%much larger than the first.  We  are continuing to run on that set in hopes
%of reducing these fluctuations.

\begin{figure}[htb]
\null
\vspace{-1.3truecm}
%\framebox[55mm]{\rule[-21mm]{0mm}{43mm}}
\includegraphics[bb = 100  600 4096 4196,
%NOTE: to get graphs made on fuji (with psplot) to be included correctly,
% must comment out "bounding-box-like" line (e.g "0 0 4096 4096 s")
% immediately after prolog in ps file
%or make it into true bounding box
width=2.5truein,height=2.3truein]{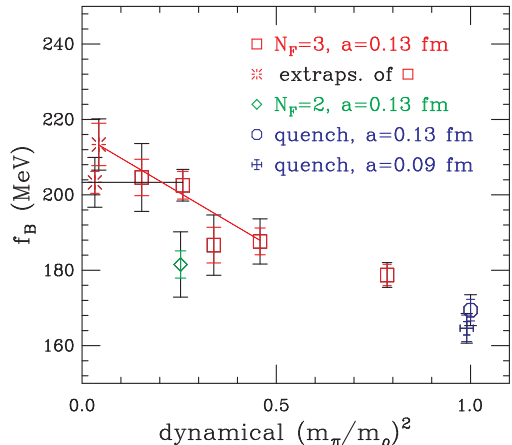}
\vskip -.1truein
\caption{$f_B$ versus 
$(m_\pi/m_\rho)^2$ of the dynamical light quarks. The scale is set by $m_\rho$, 
with valence quarks extrapolated to the physical point. } 
\label{fig:fb}
\vspace{-0.3truein}
\end{figure}

Preliminary data for $f_B$ with the scale set by
$m_\rho$ is shown in Fig.~\ref{fig:fb}.  
%uses linear chiral fits for the heavy-light masses
%and decay constants, 
There is clear evidence for an increase in $f_B$ as the dynamical quark mass is
decreased.  There is also an increase
% although not as significant
 in the $N_F\!=\!3$ results over those with $N_F\!=\!2$.  Note 
that the quenched results at
the two lattice spacings are consistent; they differ by $\ltwid 3\%$.

\begin{figure}[htb]
\null
\vspace{-1.3truecm}
%\framebox[55mm]{\rule[-21mm]{0mm}{43mm}}
%NOTE: to get graphs made on fuji (with psplot) to be included correctly,
% must comment out "bounding-box-like" line (e.g "0 0 4096 4096 s")
% immediately after prolog in ps file
%or make it into true bounding box
\includegraphics[bb = 100  600 4096 4196,
width=2.5truein,height=2.3truein]{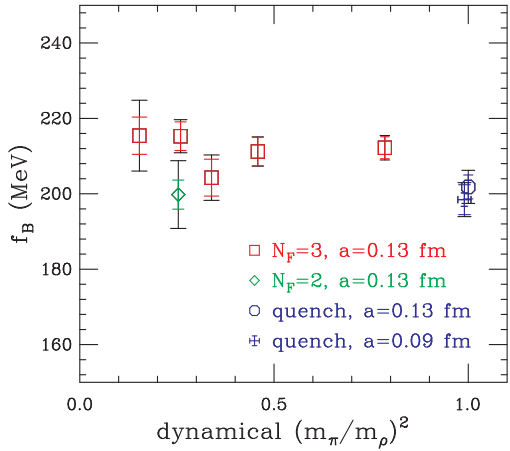}
\caption{Same as Fig.~\protect{\ref{fig:fb}}, but with scale set by
$r_1=0.35$ fm.}
\label{fig:fb_r1}
\vspace{-0.3truein}
\end{figure}

Figure \ref{fig:fb_r1} shows $f_B$ with the scale set by $r_1=0.35$ fm
(the  phenomenological
error in this value of $r_1$
is not included).
%rather than $m_\rho$.  
There is now little if any dependence on the dynamical quark mass:
Holding $r_1$ fixed forces the short distance potentials to match quite closely,
which in turn keeps $f_B$, the ``wave function at the origin,'' more or less fixed.
%It also points up an important point: the question ``what is the quenching effect
%on $f_B$?'' is by itself meaningless.  One must specify how the quenched scale is fixed.
%It is gratifying to 
In the physical case (three dynamical quarks extrapolated to
physical masses) the values of $f_B$ in Figs.~\ref{fig:fb}
and \ref{fig:fb_r1} are consistent.

\begin{figure}[htb]
\null
\vspace{-1.3truecm}
%\framebox[55mm]{\rule[-21mm]{0mm}{43mm}}
%NOTE: to get graphs made on fuji (with psplot) to be included correctly,
% must comment out "bounding-box-like" line (e.g "0 0 4096 4096 s")
% immediately after prolog in ps file
%or make it into true bounding box
\includegraphics[bb = 100  600 4096 4196,
width=2.5truein,height=2.3truein]{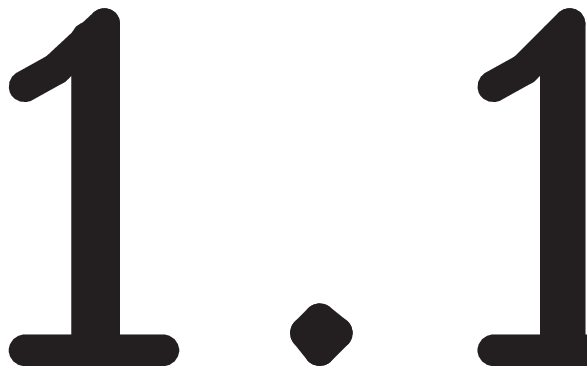}
\caption{Same as Fig.~\protect{\ref{fig:fb}}, but for \fbsofb.}
\label{fig:fbsofb}
\vspace{-0.3truein}
\end{figure}

The ratio \fbsofb\ is shown in Fig.~\ref{fig:fbsofb}.  No obvious dependence
on the number or masses of dynamical quarks is apparent.  We therefore fit
all the data to a constant.

For the moment, we assume that the scaling errors in $f_B$ or \fbsofb\ are no larger than
the difference in our quenched results at 0.13 fm and 0.09 fm.
With this assumption (which will be tested in the
coming year) we arrive at the following {\sl preliminary} results:
$f_B/f_B^{\rm quench}=1.23(4)(6)$ and  $\efbsofb=1.18(1)({}^{+4}_{-1})$,
where the first error is statistical and the second is systematic,
including discretization and chiral extrapolation (valence and dynamical) errors, 
the uncertainties in $\kappa_s$ (for \fbsofb), and
a rough estimate of perturbative errors.  A direct determination of $f_B$ itself with
three dynamical flavors must await the calculation of the renormalization factors.

We thank NPACI, The Alliance, PSC, NERSC, LANL, Indiana University, 
and Washington University (CSPC) for computing resources.
This work was supported in part by the DOE and NSF.

\end{document}